**Broken symmetries and excitation spectra of interacting electrons in partially filled Landau levels**

**Authors:** Gelareh Farahi[1†], Cheng-Li Chiu[1†], Xiaomeng Liu[1†], Zlatko Papic[2], Kenji Watanabe[3], Takashi Taniguchi[4], Michael P. Zaletel[5], Ali Yazdani[1*]

**Affiliations:**

[1] Joesph Henry Laboratories and Department of Physics, Princeton University, Princeton, NJ 08544, USA

[2] School of Physics and Astronomy, University of Leeds, Leeds LS2 9JT, UK

[3] Research Center for Functional Materials, National Institute for Materials Science, 1-1 Namiki, Tsukuba 305-0044, Japan

[4] International Center for Materials Nanoarchitectonics, National Institute for Materials Science, 1-1 Namiki, Tsukuba 305-0044, Japan

[5] Department of Physics, University of California at Berkeley, Berkeley, CA 94720, USA

*Correspondence to: yazdani@princeton.edu

† These authors contributed equally to this work

**Interacting electrons in flat bands give rise to a variety of quantum phases. One fundamental aspect of such states is the ordering of the various flavours – such as spin or valley – that the electrons can undergo and the excitation spectrum of the broken symmetry states that they form. These properties cannot be probed directly with electrical transport measurements. The zeroth Landau level of monolayer graphene with four-fold spin-valley degeneracy is a model system for such investigations, but the nature of its broken symmetry states – particularly at partial fillings – is still not understood. Here we demonstrate a non-invasive spectroscopic technique with a scanning tunneling microscope and use it to perform measurements of the valley polarization of the electronic wavefunctions and their excitation spectrum in the partially filled zeroth Landau level of graphene. We can extract information such as the strength of Haldane pseudopotentials that characterize the repulsive interactions underlying the fractional quantum states. Our experiments also demonstrate that fractional quantum Hall phases are built upon broken symmetry states that persist at partial filling. Our experimental approach**

**quantifies the valley phase diagram of the partially filled Landau level as a model flat band platform which is applicable to other graphene-based electronic systems.**

When kinetic energy is quenched, interactions dominate electrons' behavior. Thus, electronic bands with flat energy-momentum dispersion are fertile playgrounds for emergent quantum phenomena. A paradigm of flat bands is Landau levels formed in two-dimensional (2D) electron systems under a strong magnetic field[1,2]. When the LL filling factor $\nu$ ($\nu = 2\pi n l_B^2$, where $l_B = \sqrt{\hbar/eB}$ is the magnetic length and n the carrier density) is an integer, Coulomb interactions give rise to quantum Hall ferromagnetism[3,4], where certain isospins (such as spin or valley) or their coherent superposition within a LL are fully occupied. At partial fillings of Landau levels, a rich variety of strongly correlated electron states, including broken symmetry phases [5] [6–9] and fractional quantum Hall (FQH) states[2,10,11], emerge due to many-body effects. The emergence of correlated phases is driven by a set of discrete energy levels, known as Haldane pseudopotentials ($V_m$) [12], which represent the energy cost for forming electron pairs with relative angular momentum m in partially filled Landau levels, where inter-electronic spacing is approximately quantized in units of $\sqrt{m}l_B$. While the pseudopotentials provide a key to the understanding of the energetic competition between correlated phases, their direct measurement has so far been limited to GaAs heterostructures [13,14]. In the zeroth LL (ZLL) of monolayer graphene, which has four-fold spin and valley degeneracy, early efforts focused on properties at integer fillings, with recent STM studies[15–17] advancing the understanding of half-filled ZLL at charge neutrality by visualizing an intervalley coherent (IVC-Kekulé) ground state. However, less is understood about the nature of symmetry breaking at partial fillings of LL, where multiple orders, including single and multicomponent FQH states[10] compete for the ground state. Spin-polarized tunneling spectroscopy measurements of Landau levels in GaAs heterostructure devices[18] have characterized spin polarization of various phases; however, in graphene's ZLL the valley degree of freedom in addition to spin texture makes the possible broken symmetry states more complex. Magnon transport experiments[19,20] on graphene probe the spin excitation spectrum across the phase diagram, but they are unable to probe valley polarization. Taking advantage of the correspondence between occupation of valley (K, K') and sublattice (A,B) polarization, here we apply microscopy and spectroscopy with the STM to determine the nature of the ground state as well as the valley texture of the quasi-particles excitation in ZLL of graphene, as function of filling. In this work, we go beyond all previous studies and identify the conditions under which spectroscopic measurements with the STM are non-invasive for fragile partially filled LL states, in which gaps are an order of magnitude smaller than integer fillings. This is confirmed by our ability to resolve Haldane pseudopotential energies for graphene in our spectroscopic measurements that are in excellent agreement with exact diagonalization calculations. Our data shows evidence for transitions between spin and valley polarized, valley unpolarized, and an intervalley coherent (IVC) states as function of filling in the ZLL. The appearance of FQH states, the presence of which we detect by their spectroscopic gaps, do not alter valley polarization at partial fillings thereby demonstrating that FQH states inherit the broken spin/valley symmetry of the neighboring states.

For our studies, we prepare ultra clean monolayer graphene devices (see methods in SI), with a hexagonal Boron Nitride (hBN) substrate on top of a graphite back gate and examine them in a homebuilt STM system as a function of gating and magnetic field (Fig. 1a-b). With hBN on only one side of the monolayer graphene, the Coulomb interaction in our samples is enhanced as compared to the hBN-encapsulated graphene devices examined in transport studies. As in our previous study[15] and given the absence of broken symmetry gaps in previous STM works[16,21–31] we have found that preparation of sharp STM tips on a Cu surface prior to measurements on graphene results in no detectable tip-gating effect, as judged by the pinning of the ZLL to the Fermi level in the dI/dV spectroscopic measurements in the tip-sample bias versus gate voltage plots (Fig. 1b-c of ref. 15, for example). However, even in the absence of such gating effects, we have discovered that the tip may still modify the spectroscopic measurements and understanding of these phenomena is critical to measurements of the intrinsic properties of this low density 2D electronic system. Figure 1c and f shows spectroscopic measurements of the ZLL with two different tips, in which we see the expected Coulomb gap near the Fermi energy, gaps due to the broken symmetry states forming at filling $\nu = -1, 0, 1$, gaps at some of the FQH states (see below) and the single particle gaps at $\nu = -2, 2$. However, the tips display distinct spectroscopic features, especially at partial fillings. Measurements with such tips often show checkmark-like features on either electron-like or hole-like excitation near gapped states reminiscent of Coulomb blockade features in transport studies of gated devices[32].

To understand how the STM tips may be influencing our measurements, we utilize exact diagonalization techniques to calculate the excitation spectrum for a 2D electron gas in a magnetic field properly accounting for the isospin degeneracy under the assumption that the ground state is spin and valley polarized (see SI for details of calculation). Such calculations should accurately capture the experimental setting between $\nu = $ -2 to -1, where we are filling one flavor out of four in the ZLL. We can accurately match the spectroscopic measurements for the two different type of tips in Figure 1c and f with theoretical calculations (figure 1d and g), between $\nu = $ -2 to -1, provided we include in the calculations a local potential (on the magnetic length $l_B$) to model the influence of the STM tip. For the tips in figure 1, we find that including a negative (-10 meV) or positive (+10 meV) amplitude local potential gives excellent comparisons between calculations and measurements in the range $\nu = $ -2 to -1. The model calculation not only captures the Coulomb gap near the Fermi energy ($E_c = 0.62 \, e^2/\epsilon l_B$ [17]) but more importantly, the details of the density dependence of the occupied and unoccupied density of states in the spectra for different tips at partial filling $\nu = $ -2 to -1. Intuitively, we can understand the features in the spectra are influenced by the STM tips trapping either an electron or hole-like quasiparticle underneath the tip (Fig. 1e and i), the strength of which depends on the details of the impurity-like potential induced by the tip on the sample.

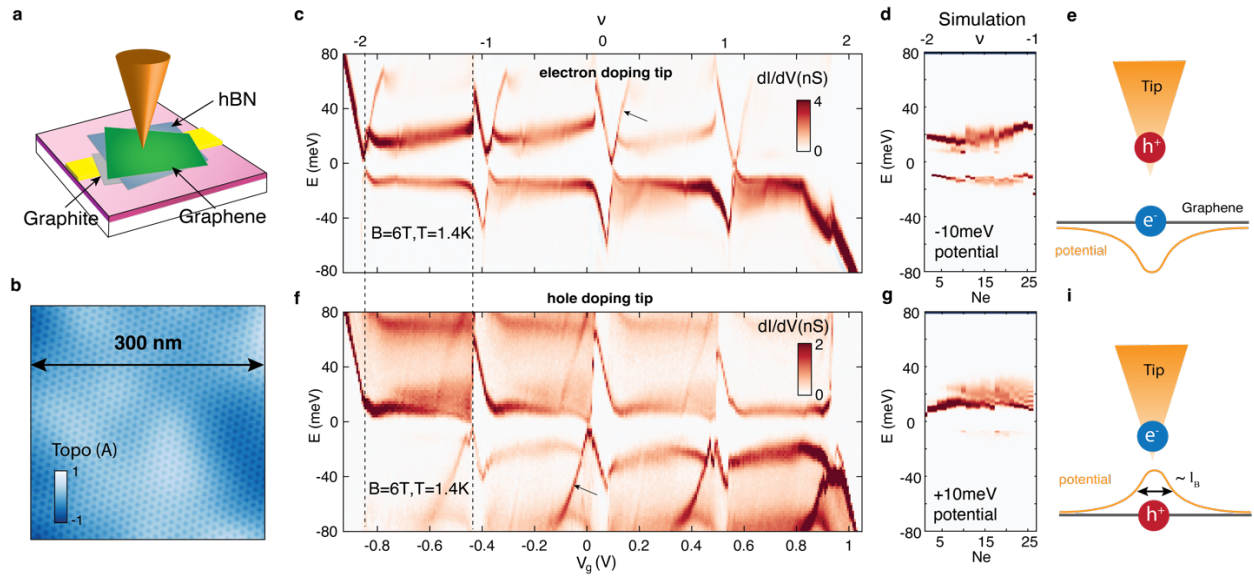

**Figure 1 Influence of the STM tip on LL spectra.** (a) Device architecture, (b) A topographic map showing the moiré pattern from the hBN substrate. (c), (d), spectroscopy of the ZLL with an electron and hole doping tip respectively, example checkmarks near $\nu = 0$ gap have been indicated with arrows. (e), (f) exact diagonalization calculation including a negative and positive impurity potential under the STM tip for a single flavor LL, capturing the -2< $\nu$ <-1 range in the experiments. (g), (h) schematic of the tip-induced impurity potential for an electron and hole doping tip respectively.

In addition to tips that induce "positive" or "negative" potentials, we have found that measurements with a third category of tips, which show neither of the checkmark-like features of the other tip-types but instead clear "sash"-like features near the Fermi energy (Figure 2a). This category of tips, which we call "neutral" tips, best reflect the intrinsic density of states as they show excellent comparison with the exact diagonalization calculation (between $v = -2$ and -1) without any perturbing potential (figure 2d). The sash-like features measured with neutral tips that are reproduced in our exact diagonalization calculations have been previously reported in tunneling spectroscopic measurements of the first LL in GaAs heterostructures and can be theoretical understood as signatures of the Haldane pseudopotentials seen in GaAs heterostructures [13,14] whose calculation has been generalized for monolayer graphene with four-fold isospin degeneracy [33]. Detailed analysis of the features in the spectra (Figure 2e and g) can be used to extract the energy scale of the "Haldane sash" features as a function of the electron density and compare it to those from exact diagonalization calculations, as shown in Figure 2f. We find that the energy gap in the electron excitation spectra ($\Delta_{sash}$), extracted by fitting a Lorentzian to the excitation peaks, in the density ranges $v = -2$ and -1.5 and $= -1$ and -0.5 have magnitude and density dependence in agreement with the exact diagonalization calculations (figure 2f), giving us confidence that our measurements are probing the intrinsic properties of interacting electrons in the ZLL. The absence of Haldane sashes in "positive" or "negative" potential tips can be understood if we consider that the tip's impurity potential to lift the orbital degeneracy of LL locally, thereby eliminating the strong interaction between the zero angular momentum state underneath the tip and other orbital states. It is noteworthy that despite the quantitative agreement in the extracted gaps from theory and experiment, the correspondence in spectral weight is often only qualitative, since the integrated density of states which is in principle proportional to the tunneling current in our measurements does not always follow the sum rule as function of filling (see SI Fig. 5). The sum rule is obtained only under the assumption that STM conductance is proportional to the electronic spectral function under the tip [34]; however, the possibility that STM tunneling perturbs the quantum Hall fluid or other processes may also need to be considered. For example, phonon-mediated inelastic tunneling processes which are commonly seen in STM measurements of graphene, and have not been included in calculations, can modify the spectral density of the STM spectra (SI Fig 5.a,b). Another complex behavior is that near $v = \pm 2$ where we sometimes observed Haldane sash behavior approaching the filled LL, where the sample is entering the insulating regime and tip-gating effects for some tips (Fig. 2a, also see SI Fig. 5) may influence the local doping.

Using neutral tips, we also characterize the gaps in the tunneling density of states as function of electron filling, revealing both the gaps associated with broken symmetry states at $\nu = -1, 0, 1$, as well as gaps of FQH states at $\nu = \pm\frac{5}{3}, \pm\frac{2}{3}$, and $\pm\frac{1}{3}$. We extract the spectroscopic gaps by measuring the gate voltage interval in which the sample is in the incompressible FQH state, which can be determined from the high-resolution spectroscopic data in figure 2b. Assuming the chemical potential $\mu = (N - CV_g)/C$, where N is the total of charge in the system and C is the device capacitance, within a gapped state carrier density is constant and thus $|\Delta\mu| = |\Delta V_g|$ and spectroscopic features follow |dVg/dE| =1. As schematically depicted in figure 2c, the Coulomb gapped spectra is modified by the jump in $\mu$ and $\Delta V_g$ is thus a direct measure of the thermodynamic gap (Δ) of the incompressible FQH state. The gaps at ⅓ fillings determined using this technique (see Fig. 2b) are all about 7 meV. After adjusting for the different dielectric environments and different quasi-particles involved in the transport and STM measurements, gaps measured with STM are comparable to thermodynamic gaps but clearly larger than transport measurements[35,36](see SI for a quantitative comparison). This difference with transport measurements is partially expected due to charge fractionalization: the transport gap measures the energy of well-separated quasiparticles of fractional charge e*, while STM spectroscopy measures the energy of q = (e/e*) such quasiparticles injected directly under the tip. Furthermore, averaging out the gap for macroscopic areas of the sample in transport and capacitance measurements are more susceptible to disorder-induced broadening, which can soften the gap while local spectroscopy is less prone to bulk impurities. We can use another approximate method to detect the chemical potential jumps by extracting the energy range in which the current is less than a threshold value |I|=0.5pA (See SI Fig. 6 a and c). These jumps which resemble the Jain sequence seen in Corbino transport[36] reveal even higher order fractional states such as $\nu = \pm\frac{2}{5}$. While we can detect chemical potential jumps at higher order fractional quantum Hall states, rigorous determination of the gap size at these fillings is not feasible from spectroscopy due to broadening effects.

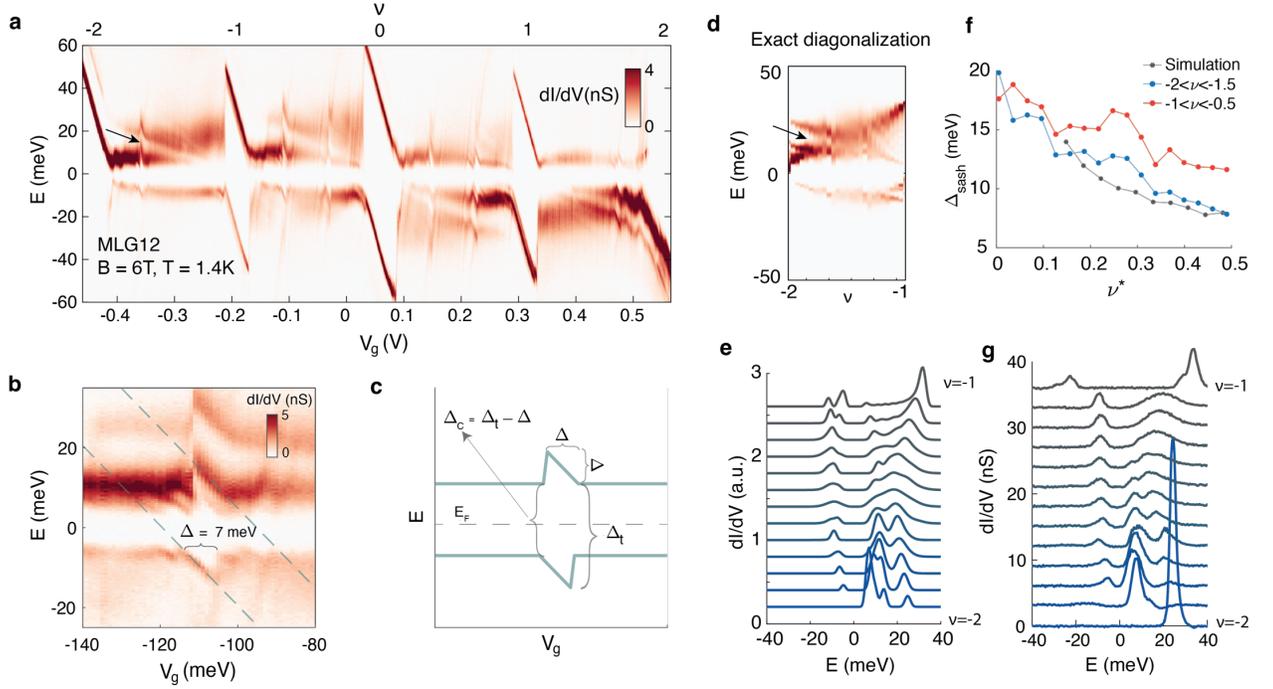

**Figure 2 Point spectroscopy of the ZLL with a charge neutral tip.** (a) False color map of gate-tunable STS showing the ZLL excitation spectra in the $-2 < \nu < 2$ filling range, arrow points to Haldane sash feature in the $-2 < \nu < -1$ (b) Zoomed spectra near $\nu = -2/3$. The FQH gap is measured using the gate range withing the gap. Dashed lines trace a $|dV_g/dE| = 1$ slope shifted in y axis to align with data. (c) Diagram showing peak positions in STS dat in green and the relationship between the tunneling gap ($\Delta_t$), thermodynamic gap ($\Delta$), and the Coulomb gap ($\Delta_c$). (d) Density of states obtained from exact diagonalization calculation for a single flavor quantum Hall system. (e) Linecuts from (d) showing the spectra at selected fillings (f) Extracted gap using Lorentzian fits to the electron excitation peaks forming the Haldane sash features in the $-2 < \nu < -1$ range (blue) and the $-1 < \nu < 0$ range (red). Similar gap extracted from exact diagonalization simulation is shown in gray. (g) Linecuts from (a), showing spectral features at constant fillings for comparison with theory (d).

Having established our ability to perform measurements of intrinsic spectroscopic properties of graphene with neutral tips, we proceed to combine these measurements with spectroscopic mapping to identify the valley texture of electronic wavefunctions and excitation spectrum of the ZLL at all fillings. In these experiments, we measure dI/dV maps with atomic resolution over 9 nm$^2$ square areas, as a function of tip-sample bias and gate voltage, obtained with a constant tip height to avoid any influence of STM feedback on the measurements. Fourier analysis of such maps (examples in Fig. 3a-c) can be used to extract information about valley polarization and intervalley coherence from intensity of the graphene sub-lattice Fourier peaks (complex wavevectors $G_0$, $G_1$, $G_2$) or when there are additional peaks (complex wavevectors $K_0$, $K_1$, $K_2$) that correspond to tripling of the unit cell in the presence of intervalley coherent (IVC) states. These peaks correspond to the Kekulé pattern in the real space maps (Figure 3c). We can determine the sub-lattice polarization, Z which is related to the out of plane angle $\theta$ on the valley Bloch sphere (Figure 3g-h, where K and K' valleys correspond to $\theta = 0°$ and $180°$ respectively for each spin spices) and can be computed from the Fourier analysis of the maps $Z = cos(\theta) = -\frac{tan(\varphi/3)}{\sqrt{3}}$ where $\varphi = arg(n(G_0).n(G_1).n(G_2))$ [15]. Z describes the normalized weight difference of the wavefunction on the A or B sub lattice in real space (ranging from 1 to -1 respectively). The intervalley coherence can also be characterized by the ratio of Kekulé peaks to those of graphene sublattice peaks and can be computed from the data $\frac{A_K}{A_G} = \frac{\sum_{i=0}^{2}|K_i|}{\sum_{i=0}^{2}|G_i|}$. As shown in figure 4, we can use this analysis to characterize the energy-resolved valley texture in terms of valley polarization and coherence. In displaying the energy-resolved valley texture as a function of filling (figure 4b and c), we use a minimum average conductance per map (0.17 nS), to exclude regions in which the density of states is too weak to be analyzed.

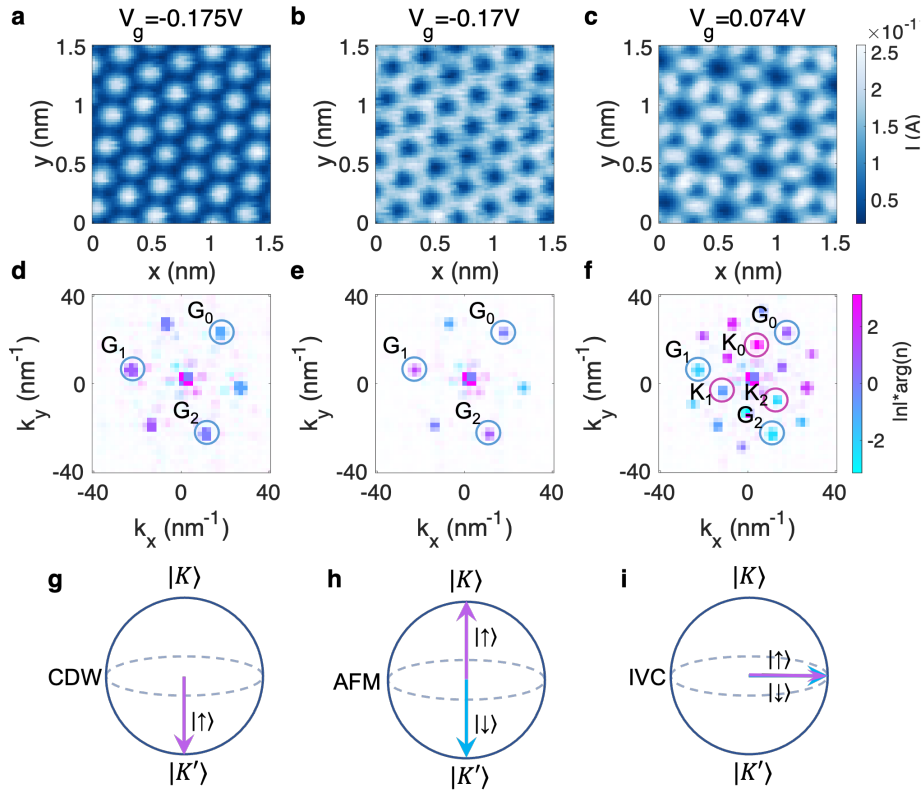

**Figure 3. Fourier analysis of constant height maps at representative fillings** (a), (b), (c) current maps of the hole excitations taken in the $\nu = -1$ gap, in $\nu = -1 + \delta$ where $\delta$ is a small number, and charge neutrality. (d), (e), (f) phase map of the Fourier transformed images, masked by the intensity of the peaks, n here is the complex Fourier amplitude. $G_i$ and $K_i$ (i = 0..2) represent the graphene and Kekulé reconstruction peaks. (g), (h), (i) Schematic of states in the valley Bloch sphere considering the two spin species marked by up and down arrows.

The data in Figure 4 has a wealth of information to understand the nature of broken symmetry states in ZLL, their excitation spectrum, and phase transitions between them, as function of filling (at 6T). Overall, both the energy-resolved (Fig. 4b) and energy-integrated (Fig. 4d) valley maps show an electron-hole symmetry, where hole-like excitation for $\nu<0$ exhibits valley polarization similar to the electron-like excitation for $\nu>0$, we see polarizations mapping into each other when switching sign of $\nu$ and sign of voltage bias. The second feature is that low energy excitations switch valley polarization (Z<0, blue, Z>0 red) across each quarter filling. Starting from $\nu=-2$, the influence of hBN substrate favors filling one sub-lattice over the other (Z<0 blue for occupied states) making a valley polarized CDW the ground state. In this regime, we can compare the valley texture of our excitation spectrum with our exact diagonalization studies, and find they have very similar features: hole-like excitations have only Z=-1 texture, while electron-like excitations show a more complicated valley dependence on energy. The details of this electron-like energy dependence can show some sample and tip variations (See SI Fig. 3), especially approaching insulating states at integer fillings, where our experiment deviates from calculations. Moreover, our energy-integrated electron-like excitation shows little or no valley polarization, a feature that is not in agreement with the theoretical expectation of a continuous increase of valley polarization from Z=0 near $\nu=-2$ to Z=⅓ at $\nu=-1$[37]. In particular, it violates the spectral weight sum rule $(2 + \nu)$ $Z_{h\text{-excitation}}$ + $(2 - \nu)$ $Z_{e\text{-excitation}}$ = 0, and must therefore be related to other processes not considered when modeling tunneling as sensing just the local density of state. At $\nu = -1$, when the broken symmetry state is gapped, we find however that Z=1/3 for electron-like excitation and Z=-1 for hole-like excitation, as one would expect for a valley polarized CDW ground state.

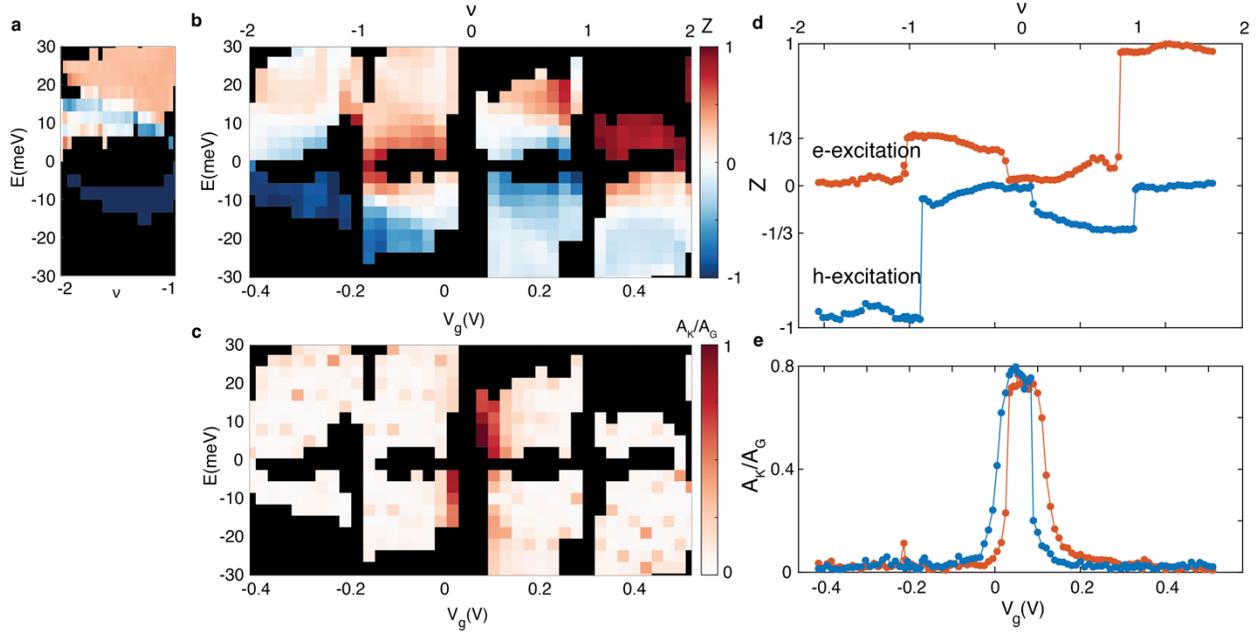

**Figure 4. Evolution of valley texture of the excitations with gate and bias** (a) Valley texture obtained in exact diagonalization calculations corresponding to filling range of -2< $\nu$ <-1 (b) extracted valley polarization (Z) from 3nmx3nm maps in the -2< $\nu$ <2 filling range of the ZLL (c) Valley coherence strength identified by the ratio of Kekulé reconstruction ($A_K$) and graphene's ($A_G$) peaks in Fourier space. (d) Valley polarization of the integrated density of states for the hole and electron excitations, using current maps with bias set outside the peaks. (e) Valley coherence of the integrated density of states, showing residual coherence outside the $\nu$ = 0 gap.

A surprising phase transition occurs as we dope the valley polarized state CDW state beyond the $\nu$ =-1 state, where there is a sudden loss of valley polarization for the occupied state, with hole-like excitation showing a jump from Z=-1 to 0. Examining the real space maps in this range (Figure 3b), we find that there is indeed density of states on both sublattices, without any intervalley coherence signal as characterized by $A_K/A_G$ in figure 4e. This is indicative that at $\nu$ =-1 the system makes a sudden transition from the CDW phase to a valley unpolarized (VUP) state. One of the proposed candidates for the spin polarization of the ground state in the filling range just below $\nu$ =-1 is an antiferromagnet (AFM) as presumed in figure 3h, while other spin textures are also possible. The Coulomb interaction at this filling overcomes the sub-lattice potential created by hBN (see also discussion below) to suddenly make electrons occupy both sublattice sites in a VUP state. Upon additional filling, we find that the VUP state smoothly evolves into a state with intervalley coherence, with A_K/A_G reaching its maximum value in the gapped $\nu$ =0 state. The smooth evolution between VUP and IVC-Kekulé states suggests the possibility that the system may enter the recently proposed canted-Kekulé state[37], in which different spin flavors have different isospin orientation on the Bloch sphere describing their interval coherence. The electron-hole symmetry of the data in figure 4, shows that the physics of $v>0$ can be understood in an analogous way to that of $v<0$.

Finally, the absence of changes in valley polarization as we transition through various FQH states at partial filling (see Fig. 4d, and SI Fig. 9 for representative maps near $v=-2/3$) indicates that these phases are built upon broken symmetry states. It is likely that between $v=-1$ to 1 the FQH state are multi-flavor, a fact that has been invoked[38] in understanding the robustness of FQH in this regime as compared to filling between $1<|v|<2$. Our observation of differences in the fractional gaps in these two regimes is consistent with this possibility.

We finally comment on the field dependence of our results (SI Fig. 1). In general, at low magnetic fields, the sublattice potential due to hBN overcomes the strength of the Coulomb interaction to drive the system valley and sublattice polarized CDW state. The transition we described above from CDW to VUP near $v=-1$ (or similar one involving hole in $v=+1$), shifts to higher filling as the strength of the magnetic field is reduced. The other significant field dependence is the presence of IVC-Kekulé states at filling close to $v=0$, which extends to a slightly wider range of filling as the magnetic field is reduced. Measurements over different devices show that as the angle of lattice of the monolayer graphene with that of the hBN is increased, the Coulomb interaction overcomes the influence of hBN potential at lower fields (see SI Fig. 2b). Given the small $\Delta_{AB}$ in the device under study (see SI Fig. 2a), 6T is a field range where Coulomb interactions are dominant and thus our measurements reflect the robust properties of the interacting electrons in the ZLL.

Looking beyond monolayer graphene, the experimental techniques we have utilized in this work to determine the valley phase diagram of ZLL can be applied to both Bernal-stacked bilayer[39] and magic angle bilayer graphene (MATBG). Study of FQH states in graphene-based systems can now be significantly advanced as local spectroscopic measurements of the thermodynamic gap may overcome limitation of macroscopic average measurements that may more sensitive to disorder. For MATBG theoretical studies[40,41] have proposed valley polarized, IVC, or more complex electronic states that have distinct atomic scale wavefunctions on the sublattice of the top layer of graphene in that system. Measurements such as those presented here on monolayer graphene would be able to discriminate between these different candidate state. Similarly, the valley texture of the excitations of MATBG, which we have measured for the ZLL quantum Hall ferromagnetic phases of graphene, could provide key information for understanding the physics of interacting electrons in MATBG[42–47]. In monolayer graphene, thus far theory has only attempted to capture the valley texture of the ZLL when only a single electron flavor is filled. Our work motivates extending such experimental measurements to MATBG, as well as future theoretical work to understand what implications the dependence of the valley texture on the electron density has for the wider isospin phase diagram.

**Acknowledgements:** We thank S. S. Hegde and I. S. Villadiego for helpful discussions.

**Funding**: This work was supported by the ARO MURI (W911NF-21-2-0147), ONR N00012-21-1-2592, Gordon and Betty Moore Foundation's EPiQS initiative grants GBMF9469, DOE-BES grant DE-FG02-07ER46419 to A.Y. Other support for the experimental work was provided by NSF-MRSEC through the Princeton Center for Complex Materials NSF-DMR-1420541, NSF-DMR-1904442. Z.P. acknowledges funding by the Leverhulme Trust Research Leadership Award RL-2019-015. M.Z. acknowledges support from the U.S. Department of Energy, Office of Science, Office of Basic Energy Sciences, Materials Sciences and Engineering Division, under Contract No. DE-AC02-05CH11231, within the van der Waals Heterostructures Program (KCWF16). K.W. and T.T. acknowledge support from the Elemental Strategy Initiative conducted by the MEXT, Japan, grant JPMXP0112101001, JSPS KAKENHI grant JP20H00354, and the CREST (JPMJCR15F3), JST.
**Author contributions:** G.F., C.C. , X.L. and A.Y. designed the experiment. G.F. and C.C fabricated the sample. G.F., C.C. and X.L. performed the measurements and analyzed the data. M.Z., Z.P. and X.L. conducted the theoretical analysis. G.F., C.C., X.L., A.Y. and M.Z. wrote the manuscript with input from all authors. **Competing interests:** Authors declare no competing interests.

**Methods:** Methods for device fabrication, STM measurements and data analysis have been included in the Supplementary Information.

**Data and materials availability:** The data that support the findings of this study are available from the corresponding author upon reasonable request.

# Supplementary Materials for
# Broken symmetries and excitation spectra of interacting electrons in partially filled Landau levels

## Materials and methods

<u>Sample preparation</u>

Samples in this work were fabricated using a mechanical transfer technique. Our pickup stamp is made from a polyvinyl alcohol (PVA) coated transparent tape, which covers a polydimethylsiloxane (PDMS) block on a glass slide handle. We first pick up monolayer graphene directly with PVA, and use graphene to pick up the bottom layers, hBN and graphite respectively to serve as the backgate. The stack was then transferred onto a SiO2/Si substrate with pre-patterned gold contacts after dissolving the PVA film with water. After fabrication, all three samples were rinsed in water, n-methyl-2-pyrrolidone (NMP), and acetone in order to dissolve polymer residue from the surface. The devices were eventually baked in UHV at 400C overnight before transferring to the STM chamber. The hBN thickness for device MLG8 and MLG12 are 58nm and 50nm, respectively.

<u>STM measurements</u>

The experiment is done in a home built UHV STM operating at T = 1.4 K. All data shown unless specified otherwise are taken at B = 6T. The measurements are performed with a tungsten tip prepared on a Cu(111) single crystal. Through controlled indentation, we shape the tip until its poke mark is confined and its spectrum features the Cu(111) surface state at the right energy. We then locate the graphene sample with a capacitance guiding technique (1).

In measurements the tip is grounded and the tunneling current is measured from the tip. Bias voltage $V_B$ is applied to the graphene sample while $V_B + V_g$ is applied to the gate to achieve a gate voltage of $V_g$ relative to the graphene sample. Differential tunneling conductances shown in this study are measured using the lock-in method with ~0.5mV AC modulation at 4 kHz. In order to circumvent setpoint effects, all the gate-tuned spectroscopy maps in this study are measured with a constant tip height, which is set at $\nu$=1/2, $V_{set} = -0.4V$, $I_{set} = $ 1-2 nA.

STM Tip Preparation

In our tip shaping recipe, we veto tips based on three main features:

- The large-scale shape of the tip must be confined. This can be confirmed by poking the tip tens of nm into Cu(111) and scanning a large ( 300x300nm$^2$) area that is centered around the poke mark. If we see a single protrusion at the location of poking that is of the same size as the depth of the poke, we proceed to next step.

- The small-scale shape of the tip also must be confined. We poke the tip about 2nm deep into copper and scan a 50x50 nm$^2$ area to ensure there are not double poke marks, if the poke mark is spherical and about the same size as the poke, we check the spectrum of the tip.

- If the spectroscopy on a clean area shows Cu(111) step-like surface state at $V_B \approx$ 470meV, and is otherwise featureless, we proceed to the measurement. Otherwise, we repeat steps (a) and (b) until criteria (c) is achieved.

It is noteworthy that the ultimate state of the tip is confirmed after tunneling into the device and making sure we see Haldane sashes in our data.

Numerical calculation of the LDOS

Numerical simulations of the LDOS in the presence of a tip-induced potential was performed using the standard formalism for many electrons projected to the $N = 0$ Landau level on the surface of a sphere (2). The symmetries of the electron system – the total $z$-projection of both spin and angular momentum of the electrons, as well as there spin and valley polarization – were explicitly resolved. The effective interaction between the electrons is modeled according to Ref. (3), and it includes the dielectric constant (for one-sided hBN) $\epsilon_{\text{hBN}} = (1 + \sqrt{\epsilon^\| \epsilon^\perp})/2 \approx 2.5$, with $\epsilon^\perp = 3.0$ and $\epsilon^\| = 5.33$, as well as the screening by the filled Dirac sea at the RPA level (4). The screening by the gate at distance $z = 50$nm below the 2DEG is weak, as this distance corresponds to $z =\sim 6\ell_B$ away from the sample. We fix the magnetic field to $B = 6$T, which is assumed to be strong enough to fully spin-polarize the ground state in the range $\leq -2\nu \leq -1$ of interest. Finally, the tip-induced impurity potential is modelled by assuming the tip is a uniformly charged sphere of radius $R$ whose surface is 1nm above the 2DEG. Because the tip potential is screened by the gate below the 2DEG, the potential produced at the 2DEG is

$$V_{\text{imp}}(r) = v\frac{(2hR + R^2)}{2h}\left(\frac{1}{\sqrt{r^2 + (h + R - z)^2}} - \frac{1}{\sqrt{r^2 + (h + R + z)^2}}\right), \quad (S1)$$

with $R = 5$nm the radius of the tip, $z = 50$nm is the sample-gate distance, and $h = 51$nm is the gate-tip distance. The overall magnitude of the impurity potential is taken to be $v = \pm 10$meV.

We evaluate the LDOS by computing the spectral functions (5, 6). Let us first ignore for a moment the spin and valley degree of freedom. The spectral functions are

$$A_+(\epsilon) = \sum_E \delta(\epsilon - (E - E_N)) |\langle N+1, E|\psi^\dagger(0)|N\rangle|^2, \quad (S2)$$

$$A_-(\epsilon) = \sum_E \delta(\epsilon + (E - E_N)) |\langle N-1, E|\psi(0)|N\rangle|^2, \quad (S3)$$

where $E_N$ is the energy of the ground state $|N\rangle$ at particle number $N$. The difference between theory and experiment in FQH gaps is that, in experiment particle number N is constant and the gate axis represents the chemical potential shift, whereas in these calculations particles are being continously added to the system. So the x-axis in theory and experiments are not equivalent in the gaps.

Note that the Hamiltonian used to evaluate the eigenstates $|N\rangle$ now includes the perturbation from the tip potential in addition to the usual Landau-level projected Coulomb interaction. If $\Delta_+ = E_{N+1} - E_N$ and $\Delta_- = E_{N-1} - E_N$, the thermodynamic charge gap is $\Delta = \Delta_+ + \Delta_-$. The chemical potential is defined by $\Delta_+ - \mu(N) = \Delta_- + \mu(N)$, or $\mu(N) = (\Delta_+ - \Delta_-)/2$. $A_+$ has support for $\epsilon > \Delta_+$ and $A_-$ has support for $\epsilon < -\Delta_-$.

The spectral functions in Eqs. (S2)-(S3) obey a number of sum rules (6, 7). For example, the zeroth-moment sum rules give the density:

$$2\pi \ell_B^2 \int A_+(\epsilon)\, d\epsilon = \langle \nu | \psi(0) \psi^\dagger(0) | \nu \rangle = 1 - \nu, \tag{S4}$$

$$2\pi \ell_B^2 \int A_-(\epsilon)\, d\epsilon = \langle \nu | \psi^\dagger(0) \psi(0) | \nu \rangle = \nu, \tag{S5}$$

where $|\nu\rangle$ denotes the ground state at filling $\nu$.

We now address the spin and valley degree of freedom. Focusing on $-2 \leq \nu \leq -1$, we assume the fractional filling is fully spin and valley polarized by the Zeeman energy $E_Z$ and sublattice potential $\Delta_{AB} = 0.6\text{meV}$ respectively (note in the $N = 0$ LL, sublattice is locked to valley). As a result, spin and valley remain conserved quantum numbers and we may define spectral functions for each isospin separately, $A_\pm^{\tau\sigma}$. Assuming the fractional filling is polarized into $\tau = +, \sigma = \uparrow$, the bulk fillings satisfy $\nu_{+,\uparrow} = 2 + \nu$ and $\nu_{\tau\sigma} = 0$ otherwise. The $A_\pm^{+\uparrow}$ spectral functions can then be computed within a single-isospin model. For the unfilled spectral functions, exact diagonalization must be performed keeping the Hilbert space of *two* isospins; the majority isospin $(+ \uparrow)$, and that of the injected / removed minority isospin.

Explicit evaluation of Eqs. (S2)-(S3) is impractical due to a sum over (in principle, all) eigenstates $|E, N \pm 1\rangle$ in the spectrum. Consequently, we used a Kernel Polymonial Method (KPM) (8) which allows to iteratively evaluate LDOS by applying a Chebyshev expansion. We use KPM expansion into a Chebyshev basis of size $\sim 100$ with the Jackson kernel from Ref. (8).

Extracting valley polarization and intervalley coherence phase from Fourier analysis of conductance maps

Focusing on one spin species (or when assuming a spin-singlet state in which the two spins are equivalent), the IQHE state is described by a spinor $(\psi_+, \psi_-) = (\cos(\theta/2), \sin(\theta/2)e^{i\phi})$ in the valley space. $\phi$ is the phase of the intervalley coherent (IVC) order while $\theta$ describes the degree of valley polarization, which we may also express as $Z = |\psi_+|^2 - |\psi_-|^2$. We would like to extract these angles from STM images of the occupied/empty density, which is in proportion to the current $I(V, \mathbf{r})$ for bias-voltages just above/below the top/bottom peak. The $K$-valley orbitals are localized at sites $\mathbf{R}_{A,i} = \mathbf{R}_i + \mathbf{r}_A$, while the $K'$ sites are localized at $\mathbf{R}_{B,i} = \mathbf{R}_i + \mathbf{r}_B$. We let $w_A(\mathbf{r})$ denote the wavefunction for the $A$ orbitals, with Fourier transform $\int e^{-i\mathbf{q}\mathbf{r}} |w_A(\mathbf{r})|^2 = F(\mathbf{q})$. For $B$, $C_2$ symmetry implies $w_B(\mathbf{r}) = w_A(-\mathbf{r})$. Since sublattice and valley are locked in the $N = 0$ LL of MLG, it will be convenient to let $\tau = \pm$ denote $(A/K)$ vs $(B/K')$ together.

An electron in orbital $m$, valley $\tau$ of the $N = 0$ LL thus has an ansatz real-space wavefunction

$$\varphi_{\tau,m}(\mathbf{r}) = \sum_{\mathbf{R}_{\tau,i}} \varphi_m(\mathbf{r}) e^{i\tau \mathbf{K} \cdot \mathbf{R}_{\tau,i}} w_\tau(\mathbf{r} - \mathbf{R}_{\tau,i}) \tag{S6}$$

Here $\varphi_m(z) \sim z^m e^{-\frac{1}{4\ell_B^2}|z|^2}$ are the lowest LL wavefunctions. The density of a uniform IQHE

state can then be obtained using

$$n(\mathbf{r}) = \sum_{m,\tau,\tau'} \psi_\tau^* \psi_{\tau'} \varphi_{\tau,m}^*(\mathbf{r}) \varphi_{\tau',m}(\mathbf{r}) \tag{S7}$$

$$= \frac{1}{2\pi\ell_B^2} \sum_{\tau,\tau',i,j} \psi_\tau^* \psi_{\tau'} e^{-i\mathbf{K}\cdot(\tau\mathbf{R}_{\tau,i}-\tau'\mathbf{R}_{\tau',j})} w_\tau^*(\mathbf{r}-\mathbf{R}_{\tau,i}) w_{\tau'}(\mathbf{r}-\mathbf{R}_{\tau',j}) \tag{S8}$$

where we have used the LLL completeness relation $\sum_m |\varphi_m(\mathbf{r})|^2 = \frac{1}{2\pi\ell_B^2}$.

**Valley polarization**

We first focus on the valley-diagonal contribution to the density $\tau = \tau'$. If the orbitals are tightly localized we can restrict to $i = j$ and obtain the contribution

$$n(\mathbf{r}) \ni \sum_i \left[ |\psi_+|^2 |w_A(\mathbf{r}-\mathbf{R}_{A,i})|^2 + |\psi_-|^2 |w_B(\mathbf{r}-\mathbf{R}_{B,i})|^2 \right] \tag{S9}$$

In Fourier space at reciprocal vector $\mathbf{G}$,

$$n(\mathbf{G}) \ni \left[ F(\mathbf{G})|\psi_+|^2 e^{-i\mathbf{G}r_A} + F^*(\mathbf{G})|\psi_-|^2 e^{-i\mathbf{G}r_B} \right] \tag{S10}$$

where $F(\mathbf{q})$ is the Fourier transform (form factor) of $|w_A(\mathbf{r})|^2$.

From this expression we would like to extract the degree of valley polarization $Z = |\psi_+|^2 - |\psi_-|^2$. However, under a shift of the origin by $\mathbf{R}_0$, the result transforms as $n(\mathbf{G}) \to n(\mathbf{G})e^{-i\mathbf{G}\cdot\mathbf{R}_0}$. Since we don't a priori know where to fix the origin in a given region of the STM image, we need a way to extract order parameters in the presence of this ambiguity.

To do so, we consider products of the form $\prod_{\mathbf{q}_i} n(\mathbf{q}_i)$, where $\sum_i \mathbf{q}_i = 0$, which is thus invariant under a shift of $\mathbf{R}_0$. Let $\mathbf{G}_i = [C_3]^i \mathbf{G}_0$, $i = 0, 1, 2$, be the three $C_3$-related reciprocal vectors. The combination $\Phi = \arg n(\mathbf{G}_0)n(\mathbf{G}_1)n(\mathbf{G}_2) = 3\arg(F(\mathbf{G})|\psi_+|^2 e^{2\pi i/3} + F^*(\mathbf{G})|\psi_-|^2 e^{-2\pi i/3}) = 3\arg(|\psi_+|^2 e^{i\alpha} + |\psi_-|^2 e^{-i\alpha})$ is thus an invariant, where $\arg F(\mathbf{G})e^{2\pi i/3} = \alpha$. Without knowing $\alpha$, we can't convert this directly to $Z$. However, if we assume $w_A(\mathbf{r}) = w_A(-\mathbf{r})$ so that $F(\mathbf{G})$ is real, $\alpha = \frac{2\pi}{3}$ or $\alpha = \frac{2\pi}{3} + \pi$, and assuming the former we find $Z = -\tan(\frac{\Phi}{3})/\sqrt{3}$. By applying this method at $\nu = -1$, where we know the sublattice splitting ensures the valley-polarized state $Z = \pm 1$, we can confirm that $\alpha \approx 2\pi/3$.

**Procedure to extract Z**

In practice, we first perform Fourier transform of the real-space dI/dV maps. The complex amplitudes of the Fourier peaks corresponding to the graphene lattice are extracted as $n(\mathbf{G}_0), n(\mathbf{G}_1), n(\mathbf{G}_2)$, where $\mathbf{G}_{0,1,2}$ are three reciprocal vectors related by $C_3$ symmetry. The sublattice polarization $Z = -\tan(\arg(n(\mathbf{G}_0)n(\mathbf{G}_1)n(\mathbf{G}_2)/3))/\sqrt{3}$, in which arg denotes taking the angle of the complex amplitude.

Extraction of the tunneling current proportional to spectral weight

In addition to the presence of the sash-like feature, the spectrum weight can also reveal the tip condition. The spectrum weight is the integral over all the electron(hole)-like excited states of the ZLL and represents the ground state of the unoccupied(occupied) state. If the workfunction between the tip and sample matched (i.e. charge-neutral tip), the density underneath the tip will follow the filling factor and the spectrum weight will be proportional to the gate voltage. While if the tip possesses a large work function difference to the sample, the local density will be pinned to the maximum or minimum of a LL allowed (n= $B/\phi_0$ or =0) and the spectrum weight will be constant over one filling factor.

Extracted spectral weights ($S$) are obtained using gate dependent currents ($I_t(V_B, \nu)$) at biases set just below(above) the hole(electron) excitation peaks ($V_B$), yielding the spectral weight for integrated density of states of the ZLL in the (un)occupied levels. In order to correct for setpoint effects (since the tip height is reset at every gate), we normalize the spectral weight by multiplying the obtained currents by a reference current curve taken at a constant height (set at $\nu$=1/2, $V_{set} = -0.4V$, $I_{set} = 1$-2 nA), setting the bias to the initial bias ($V_{initial}$) where each I/V curve is measured, turning off the feedback and recording the current while the gate voltage is swept across the filling range of the ZLL. In this case the normalized spectral weight is:

$$S(\nu) = \frac{I_t(V_B, \nu) I_{ref}(V_{initial}, \nu)}{I_t(V_{initial}, \nu)} \tag{S11}$$

The spectrum weight versus gate voltage from sample MLG8 with a charge neutral tip is shown in Figure 5d. The tunneling current which represents the spectrum weight of (un)occupied state is linear with respect to filling. This indicates the electron density below the tip is following $V_g$ and is free from tip influence. Similar behavior is also observed in the sample MLG12 as shown in 5b. Although the spectrum weight slightly deviates from straight line, the fact that the spectrum weight doesn't pin to a constant value reflects the tip has little influence on the sample.

Comparison of 1/3 gap size between STM measurement and other measurements

To draw a relevant comparison of 1/3 gap size between different references and our study, we scale the 1/3 gap size with Coulomb energy $E_c = e^2/4\pi\epsilon l_B$ based on the field and the dielectric environment. $l_B$ is the magnetic length which depends on the field used in the study. $\epsilon$ is dielectric constant which depends on the device geometry. In references 35 and 36, the graphene is double-encapsulated in hBN whose $\epsilon_{\text{hBN}} = \sqrt{\epsilon^{\parallel}\epsilon^{\perp}} \approx 4\epsilon_0$. While the device in this study has one side touching hBN and the other side exposed to vacuum, so the dielectric constant $\epsilon_{\text{hBN}} = (1 + \sqrt{\epsilon^{\parallel}\epsilon^{\perp}})/2 \approx 2.5\epsilon_0$. The following table is the summary of $\Delta_{1/3}/E_c$ comparison:

|  | $\Delta_{1/3}$ | $l_B$ | $\epsilon$ | $\Delta_{1/3}/E_c$ |
| --- | --- | --- | --- | --- |
| ref 35 | $5.5 meV$ | 6.9nm | $4.0\epsilon_0$ | 0.11 |
| ref 36 | $5.9 meV^*$ | 6.9nm | $4.0\epsilon_0$ | 0.11 |
| this study | $7 meV$ | 11nm | $2.5\epsilon_0$ | 0.13 |

*we multiply $\Delta_{1/3}$ in ref 36 by three because $\Delta_{activatoin} = \frac{e*}{e}\Delta_{thermaldynamic}$

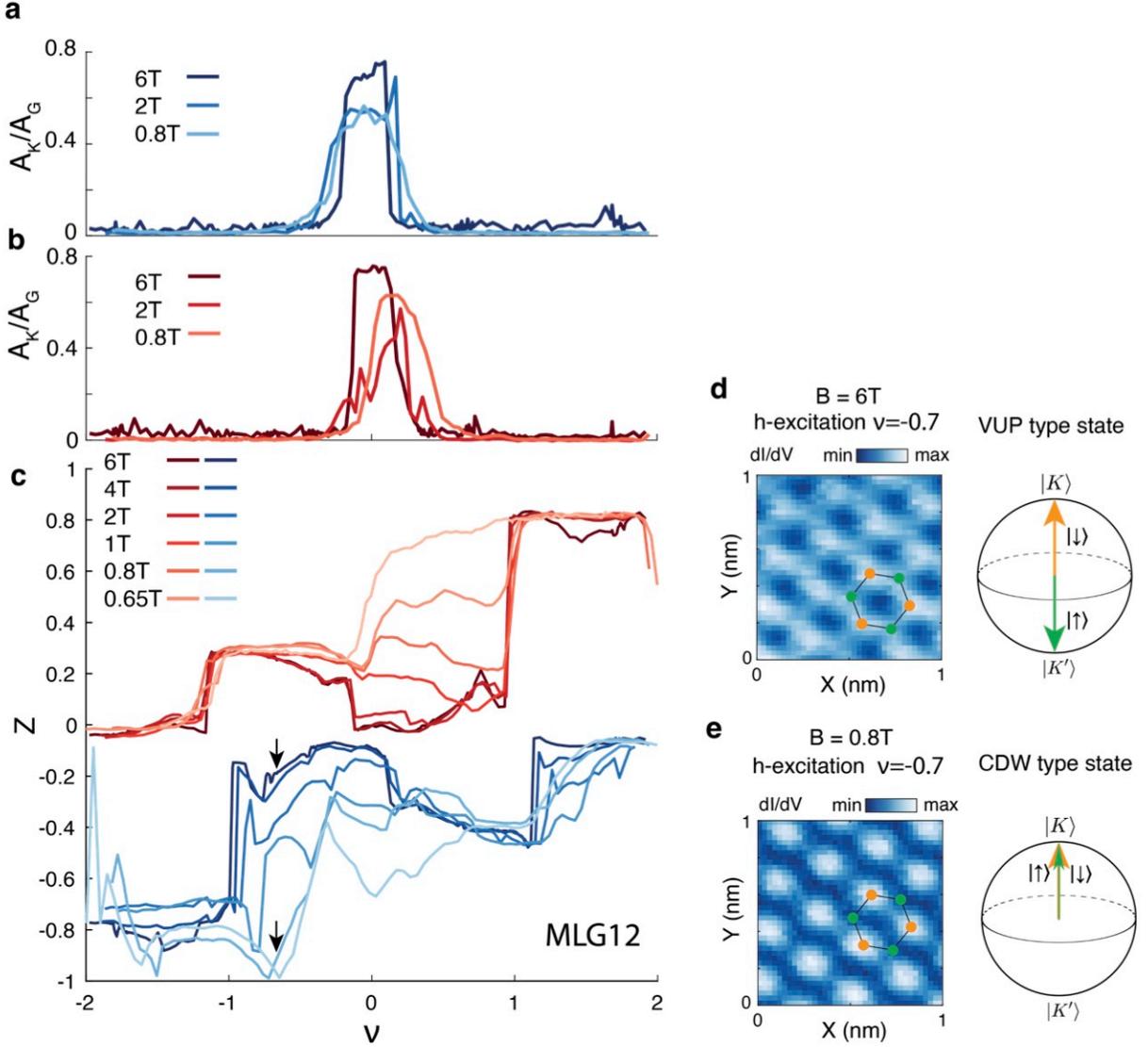

Figure 1: **Magnetic field dependence of valley texture of the ZLL. a, b** Magnetic field dependence of valley coherence in the ground state of the occupied (blue) and unoccupied (red) levels, showing a leaking of the Kekule order beyond the $\nu = 0$ gap. **c** Magnetic field dependence of valley polarization, showing a strong field dependence in the $-1 < \nu < 0$ in the ground state (blue lines, with black arrows), with the same behavior observed in the $0 < \nu < 1$ in the unoccupied levels(red lines). **d,e** Candidate Bloch vectors for the partially filled landau level in the $-1 < \nu < 0$ range are at high and low magnetic fields, respectively.

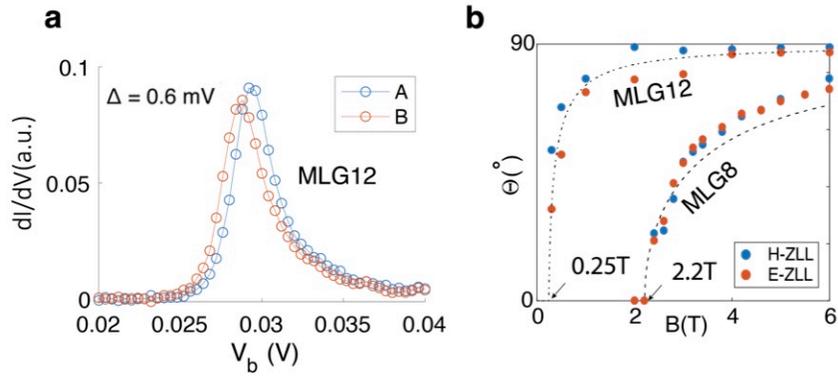

Figure 2: **Comparison of two devices with different hBN alignments a** dI/dV Spectrum in the $\nu$ = -2 gap taken on A (blue) and B(red) sites in the sample discussed in paper (MLG12), indicating $\Delta_{hBN} = 0.6$ meV obtained from Lorentzian fits. **b** Comparison of $B_c$ of MLG12 with another device MLG8, the latter with a higher hBN alignment and $\Delta_{hBN} = 6$ meV (9).

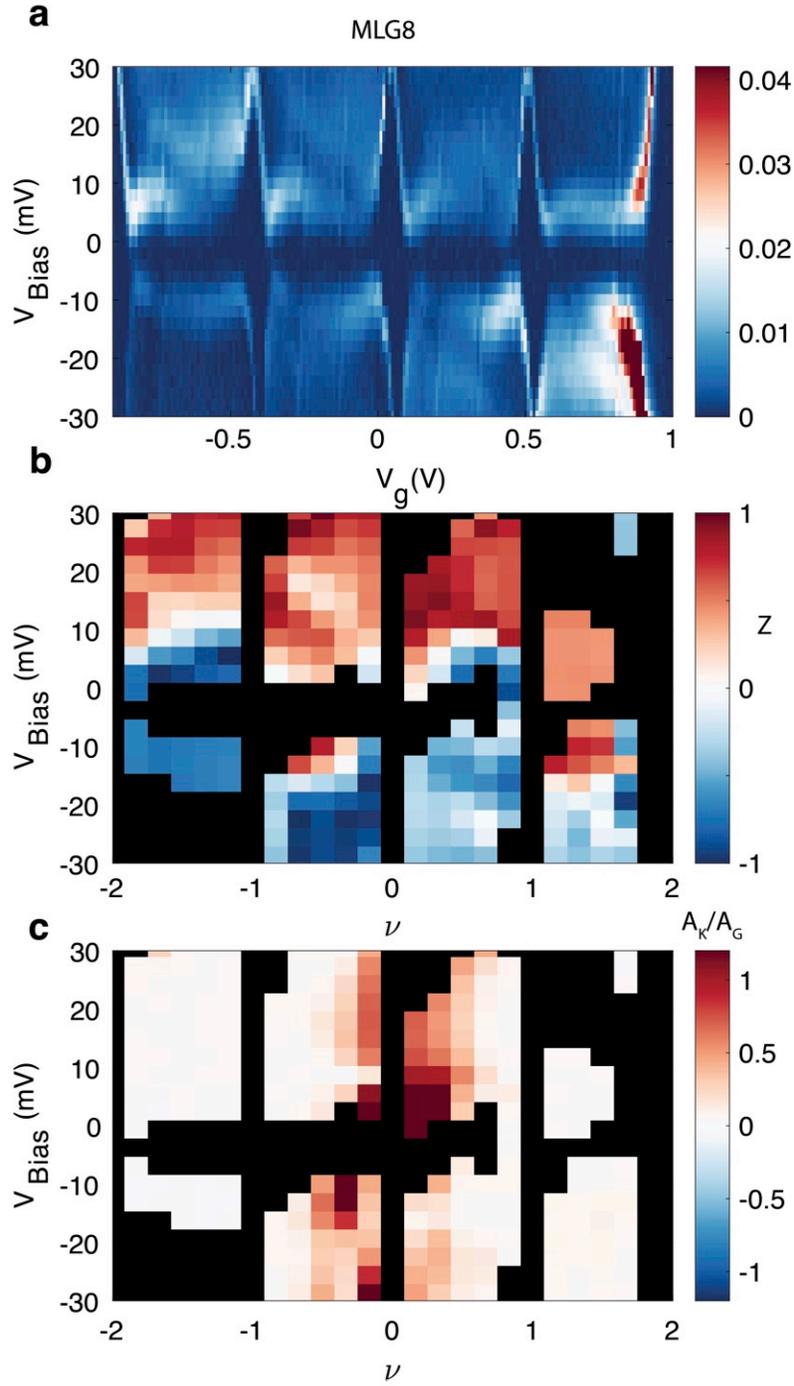

Figure 3: **Valley polarization of excitations in MLG8.** **a** dI/dV spectrum in the $-2 < \nu < 2$ range showing Haldane sashes. **b** sublattice polarizatin and **c** IVC Kekule strength. The IVC state persists in a larger filling range than that of MLG12 (see Fig. 4 in main text)

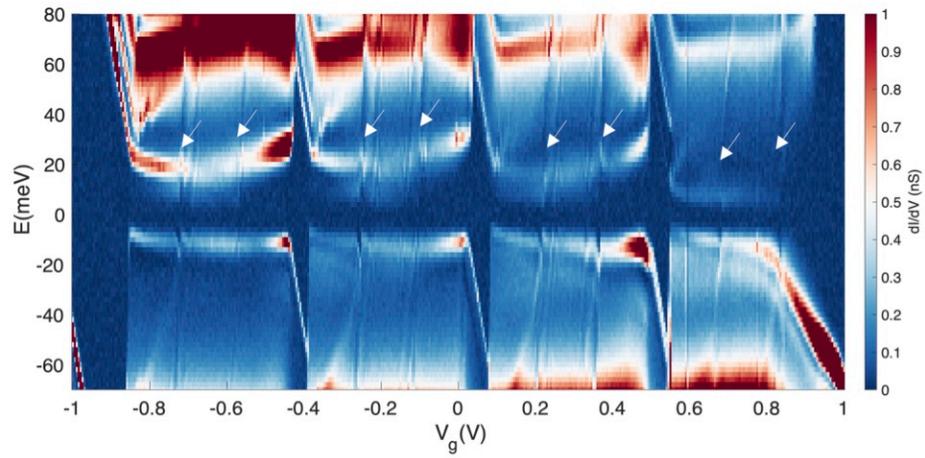

Figure 4: **Point spectroscopy performed on MLG8** arrows show the 1/3 FQH gaps.

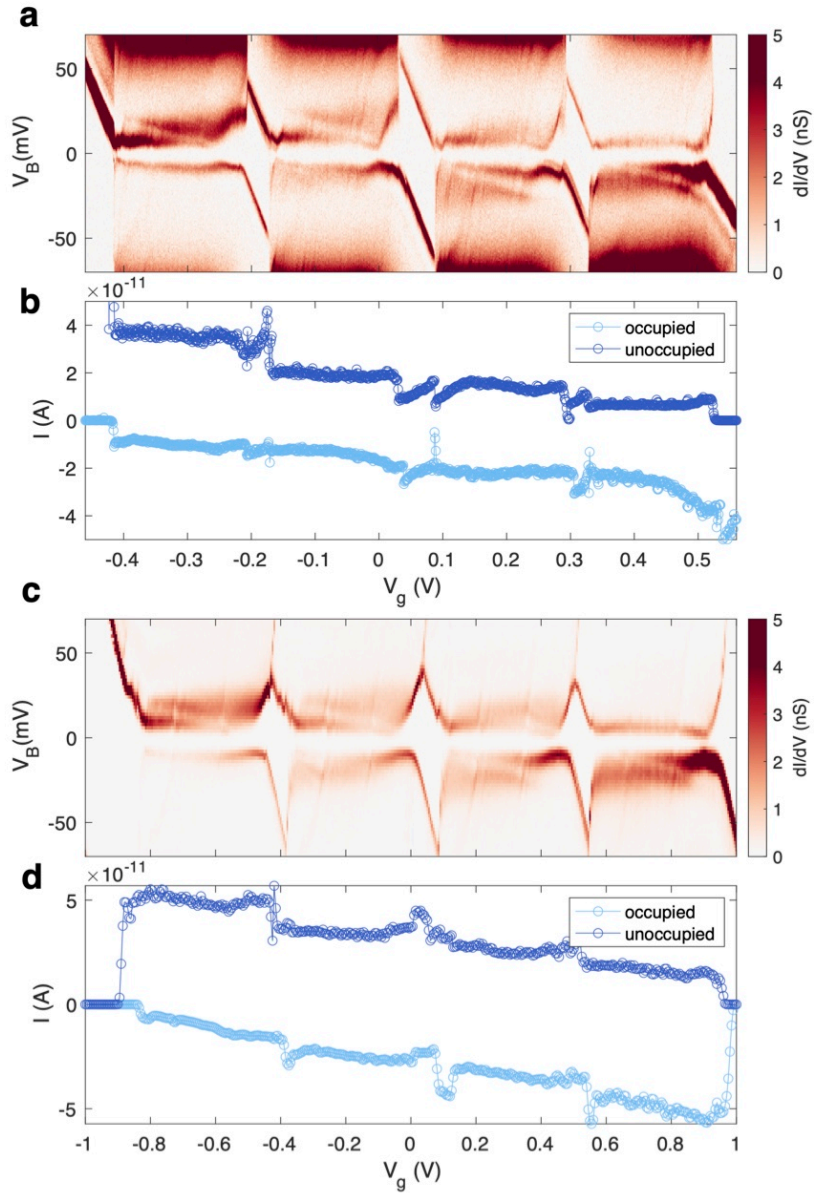

Figure 5: **Filling dependent spectral weight for two different tips. a**, **b** dI/dV and spectral weight obtained from a tip that shows a phonon gap feature, marked by enhanced density of states in ± 70 meV. **c**, **d** Similar data obtained using a tip that does not have a phonon gap, where the spectral weight shows linear dependence on filling and abides by the sum rule. Spectral weight obtained in **b**, **d** are described in SI text.

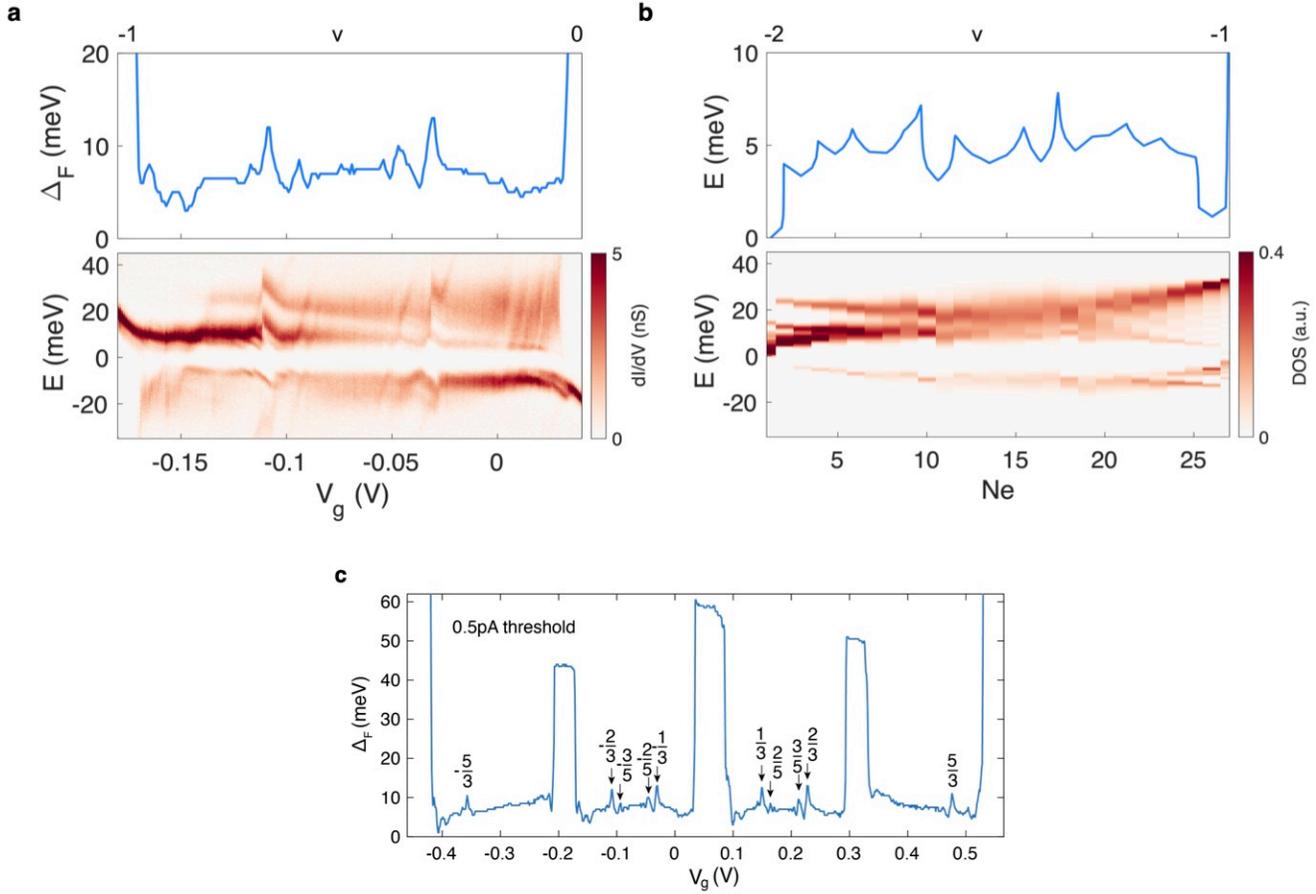

Figure 6: **Comparison of 1/3 gaps in experiment and theory at B = 6T**. **a** top panel is the extracted gaps using the same method described in main text. Bottom panel shows the spectra zoomed in the $-1 < \nu < 0$. We chose this filling range to compare with theory given the weaker 1/3 gaps in $2 < |\nu| < 1$ range. **b** Top panel shows a constant density of state contour of value of 0.2, extracted from exact diagonalization calculation shown in the bottom panel. **c** Extracted chemical potential jumps for the full filling range ($-2 < \nu < 2$) of the ZLL.

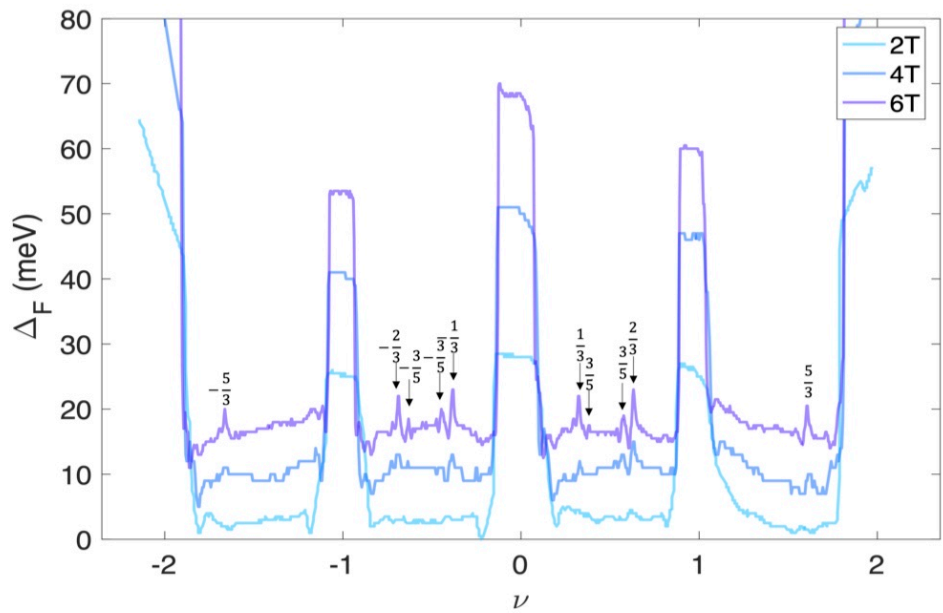

Figure 7: **Magnetic field dependence of chemical potentials.** Curves are extracted with the same method described in main text. Chemical potential jumps at FQH states are weakened at lower fields.

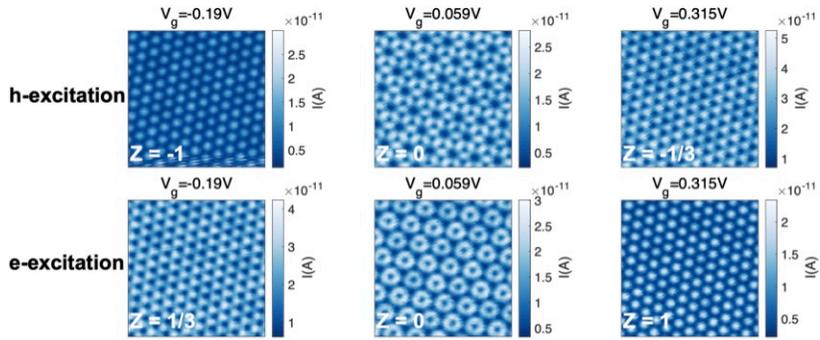

Figure 8: **Constant height 3x3 nm² current maps in summetry-breaking gaps.** Left, middle and right showing the integrated DOS for $\nu$ = -1, 0, 1 respectively. The corresponding valley polarization Z is indicated for each map.

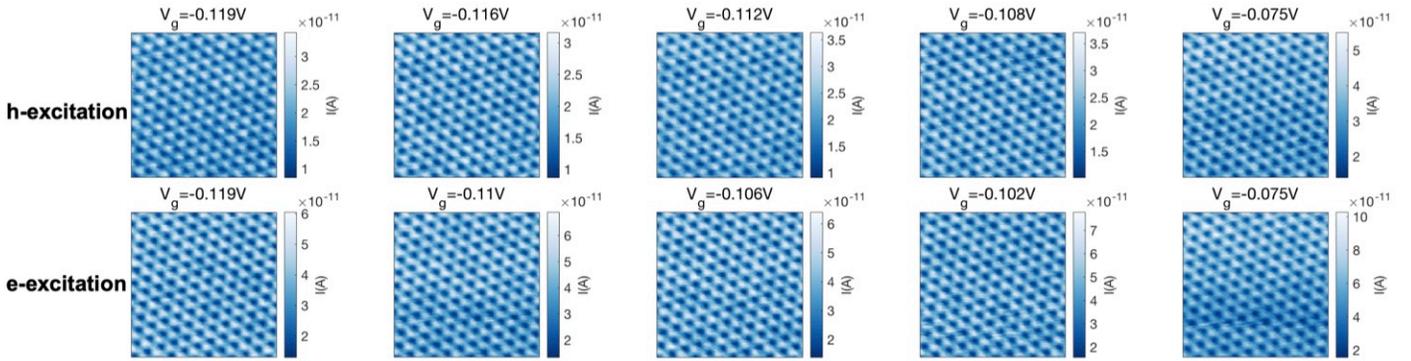

Figure 9: **Constant height 3x3 nm² current maps near $\nu = -2/3$, with $\Delta_{-2/3} = 7\text{meV}$.** Left and right panels show maps just outside the gap, three middle panels show representative maps in the gap for the hole(electron) excitation peaks in the top(bottom) panel. The small gate difference between maps in the hole and electron excitation is to incorporate the tilt in the 2/3 gap.